\documentclass{rsos}

\usepackage{hyperref}
\usepackage{color}
\usepackage{graphicx}
\usepackage{harvard}
\usepackage{ulem}

\newcommand{\vtheta}{\vec{\theta}}
\newcommand{\Nbox}{N_\mathrm{box}}
\newcommand{\Ncrit}{N_\mathrm{crit}}

\newcommand{\be}{\begin{equation}}
\newcommand{\ee}{\end{equation}}
\newcommand{\bel}[1]{\begin{equation}\label{#1}}
\newcommand{\ba}{\begin{eqnarray}}
\newcommand{\ea}{\end{eqnarray}}
\newcommand{\bal}[1]{\begin{eqnarray}\label{#1}}

\newcommand{\order}[1]{\mathcal{O}\left( #1 \right)}

\begin{document}
\title[Efficient Jump Proposals in RJMCMC]{An Efficient Interpolation
  Technique for Jump Proposals in Reversible-Jump Markov Chain Monte
  Carlo Calculations}

\date{\today}

\author{W M Farr$^{1,3}$, I Mandel$^{2,3,4}$ and D Stevens$^{1,5}$}

\address{\small{$^1$ Center for Interdisciplinary Exploration and Research in
  Astrophysics (CIERA), Department of Physics and Astronomy,
  Northwestern University, Evanston IL USA\\
$^2$ MIT Kavli Institute, Cambridge MA USA\\
$^3$ School of Physics and Astronomy, University of
  Birmingham, Birmingham UK\\
$^4$ Monash Center for Astrophysics and School of Physics and Astronomy, Monash University, Clayton, VIC Australia\\
$^5$ Department of Astronomy, The Ohio State
  University, Columbus OH USA}}

\subject{Astrostatistics, Data Analysis, Markov chain Monte Carlo}

%\eads{\mailto{w.farr@bham.ac.uk}, \mailto{imandel@star.sr.bham.ac.uk} \mailto{stevens@astronomy.ohio-state.edu}}

\corres{Ilya Mandel\\
\email{imandel@star.sr.bham.ac.uk}}

\begin{abstract}
Selection among alternative theoretical models given an observed data set is an important challenge in many areas of physics and astronomy.  Reversible-jump Markov chain Monte Carlo (RJMCMC) is an extremely powerful technique for performing Bayesian model selection, but it suffers from a fundamental difficulty: it requires jumps between model parameter spaces, but cannot efficiently explore both parameter spaces at once.  Thus, a naive jump between parameter spaces is unlikely to be accepted in the MCMC algorithm and convergence is correspondingly slow.  Here we demonstrate an interpolation technique that uses samples from single-model MCMCs to propose inter-model jumps from an approximation to the single-model posterior of the target parameter space.  The interpolation technique, based on a kD-tree data structure, is adaptive and efficient in modest dimensionality.  We show that our technique leads to improved convergence over naive jumps in an RJMCMC, and compare it to other proposals in the literature to improve the convergence of RJMCMCs.  We also demonstrate the use of the same interpolation technique as a way to construct efficient ``global'' proposal distributions for single-model MCMCs without prior knowledge of the structure of the posterior distribution, and discuss improvements that permit the method to be used in higher-dimensional spaces efficiently.
\end{abstract}

%\pacs{02.70.Tt, 95.75.Pq, 07.05.Kf}

%\submitto{\CQG}

\maketitle

\section{Introduction}

Selection among alternative theoretical models given an observed data
set is an important challenge in many areas of physics and astronomy.
In a Bayesian context, model selection involves computing the evidence
for the data given each model.  The model evidence is an integral of
the unnormalized posterior probability distribution over the model
parameter space, representing the probability of obtaining the data
set within that model.  Models with larger evidence are preferred; the
ratio of the evidences of two models is the Bayes factor between them.
The product of the Bayes factor and the ratio of prior probabilities
for the two models yields the odds ratio for the models.

There are many ways to compute model evidences.  In low-dimensional
parameter spaces, the unnormalized posterior probability can be
evaluated on a grid or lattice and the integral can be performed
directly.  For many problems or models of interest, however, the
dimensionality of the parameter space is too large to make this approach
practical, and stochastic sampling must be used.  

Markov chain Monte Carlo (MCMC) methods attempt to stochastically
produce parameter samples with density proportional to the posterior
probability distribution.  In MCMC techniques, the primary target is
an accurate estimate of the posterior distribution.  (We note that an
alternative stochastic method for exploring a model parameter space,
nested sampling \cite{Skilling:2004,Skilling:2006,Feroz:2009},
focuses on evidence computation rather than sampling the posterior
probability density functions.)  It is not straightforward to compute
the model evidence from MCMC samples.  The most direct way to estimate
the evidence for a model from MCMC samples is to compute the
harmonic-mean estimator, but this estimator of the evidence  can suffer
from infinite variance 
\cite{NewtonRaftery:1994,Chib:1995,vanHaasteren:2009,WolpertSchmidler:2012}.  MCMC
implementations with parallel tempering \cite{SwendsenWang:1986,EarlDeem:2005} allow
for evidence computation via thermodynamic integration \cite{FrielPettitt:2008}, but these can
be computationally costly.

\cite{Weinberg2009} gives a method for directly computing
the evidence integral from existing MCMC samples by using a kD-tree
data structure to decompose a parameter space into boxes containing
the MCMC sample points.  The integral is approximated as a sum over
box volumes.  This method is promising, but it is not clear in general
what statistical and systematic errors it introduces and how these are
affected by the shape of the posterior distribution which the
MCMC samples.

When the goal is model selection between several known models, only
the \emph{relative} evidence of each model is needed.  In this
circumstance, the Reversible Jump MCMC technique first introduced in
\cite{Green1995} is one of the most reliable and accurate ways to
compare the models.  Reversible Jump MCMC (RJMCMC), described more
fully in Section \ref{sec:reversible-jump}, performs a standard MCMC
in space that is an augmented union of all the model parameter
spaces.  Such an MCMC involves both intra- and inter-model jumps; the
number of MCMC samples in each model's parameter space is proportional
to that model's relative evidence in the suite of models being
compared.

Implemented naively, RJMCMC has a significant drawback: because the
chain of samples must be Markovian, only the current sample is
available to the algorithm as it is choosing the next sample.  Each
time an RJMCMC transitions between models, the information about the
choices of parameter values in the previous model is lost; subsequent
jumps into that model must ``start fresh,'' and are correspondingly
unlikely to be accepted, delaying convergence of the RJMCMC sample
chain (see Section \ref{RJMCMC} for a caveat).  
\cite{Littenberg2009} addressed this issue by proposing a new
method for producing inter-model jumps in an RJMCMC that relies on
interpolating single-model posterior distributions using a box
decomposition of parameter space.

Here we introduce an alternative technique based on a kD-tree data 
structure to construct an approximation
to each model's posterior parameter distribution.  This improved interpolation method leads to faster convergence of RJMCMC sample chains.
We draw jump proposals into the model from this approximation to its posterior.
Because jumps are proposed preferentially to locations favored by the
single-model posterior, the RJMCMC compares ``good'' locations in
parameter space across all the models, and convergence is generally
rapid.  We have successfully applied this RJMCMC technique to a 10-way
model selection among alternative mass distribution models for
black-hole X-ray binaries \cite{Farr2010}.  We also provide an example
using this method as an ``asymptotically Markovian''
\cite{terBraak2008} jump proposal in the context of a single-model,
nine-dimensional MCMC in Section \ref{sec:higherdim}.

The method of \cite{Littenberg2009} for producing inter-model jumps in
an RJMCMC relies on a box decomposition of parameter space, using
fixed-sized boxes.  The method cannot adapt to the local structure of
the posterior, and becomes asymptotically inefficient for
high-dimensional parameter spaces or highly peaked posteriors.
Meanwhile, the approximation to the posterior distribution produced by
the kD-tree is a constant-in-box interpolation of the posterior,
similar in spirit to the phase-space density interpolants produced
from N-body positions and momenta in \cite{Ascasibar2005}.  The
kD-tree interpolation is effective in parameter spaces of modest
dimensionality, and is quite space-efficient, requiring $\order{N}$
storage space and $\order{\log N}$ time to produce each proposed jump,
where $N$ is the number of samples in an MCMC over the parameter space of one model (`single-model MCMC') used to construct the interpolation.

The structure of this paper is as follows.  In Section
\ref{sec:reversible-jump} we introduce in more detail the concept of a
Reversible Jump MCMC, and describe the fundamental difficulty with a
naive jump proposal in an RJMCMC.  In Section \ref{sec:kDTree} we
introduce the kD-tree data structure used to decompose the parameter
space into boxes for interpolation.  In Section \ref{sec:efficiency}
we demonstrate the efficiency gains that are achieved from use of the
interpolated jump proposal.  In Section \ref{sec:higherdim} we give
examples of some other uses of the interpolated jump proposal that
suggest its utility in the context of a single-model MCMC.  Finally,
in Section \ref{sec:conclusion} we offer a summary and some concluding
remarks on the method.

\section{Reversible Jump MCMC}
\label{sec:reversible-jump}

Reversible jump Markov chain Monte Carlo (RJMCMC) \cite{Green1995} is
a technique for Bayesian model comparison.  Below, we give a very
brief introduction to Bayesian analysis, describe a standard MCMC, and
introduce RJMCMC.

\subsection{Bayesian analysis}

Consider an observed data set $d$ and a set of competing models for
the data, indexed by an integer $i$: $\{M_i | i = 1, 2, \ldots \}$.
Each model has some continuous parameters, $\vtheta_i$; given the
model and its parameters, we can make a prediction about the
likelihood of observing the experimental data: $L(d|\vtheta_i, M_i)$.
Within the framework of each model, Bayes' rule gives us a way to
compute the posterior probability distribution function (PDF) for the
model parameters implied by the data:
\be
  p(\vtheta_i | d, M_i) = \frac{L(d|\vtheta_i, M_i) p(\vtheta_i|M_i)}{p(d|M_i)},
\ee
where $p(\vtheta_i |d, M_i)$ is the posterior distribution for the
model parameters $\vtheta_i$ implied by the data in the context of
model $M_i$, $p(\vtheta_i|M_i)$ is the prior probability of the model
parameters that represents our beliefs before accumulating any of the
data $d$, and $p(d|M_i)$, called the evidence, is an overall
normalizing constant that ensures that $p(\vtheta_i|d,M_i)$ is
properly normalized as a probability distribution on the $\vtheta_i$.
This implies that the evidence is equal to
\bel{evidence}
  p(d|M_i) = \int_{V_i} d\vtheta_i L(d|\vtheta_i, M_i) p(\vtheta_i|M_i),
\ee
where $V_i$ is the parameter space volume in model $M_i$.  For model
comparison, we are interested in the posterior probability of a
particular model, $M_i$, given the data, $p(M_i|d)$.  Using Bayes'
rule, we see that this involves the evidence, Eq.~(\ref{evidence}):
\be
p(M_i|d) = \frac{p(d|M_i) p(M_i)}{p(d)},
\ee
where $p(M_i)$ is our a priori belief in model $M_i$ and $p(d)$ is a
normalizing constant,
\be
p(d)=\sum_i p(d|M_i) p(M_i).
\ee

When selecting among alternative models, we are interested in finding
the model with the highest posterior probability $p(M_i|d)$.  However,
attempts to directly compute the evidence by performing the
integration in Eq.~(\ref{evidence}) are generally very difficult in a
multi-dimensional, multi-modal parameter space when the likelihood has
to be evaluated numerically.  In particular, a grid-based integral
quickly becomes computationally unfeasible as the dimensionality of
$\vtheta$ exceeds a few.  The parameter space must typically be
explored in a stochastic manner before the evidence integral can be
computed.  There are several stochastic parameter-exploration
techniques focused directly on evidence computation (e.g., nested
sampling \cite{Skilling:2004,Skilling:2006} and its variant MultiNest
\cite{Feroz:2009}).  Although nested sampling can be used to compute
the posterior PDFs within each model along with the evidences for the
various models, the most common technique for computing posterior PDFs
in the context of a model is the Markov chain Monte Carlo, which we
now describe.

\subsection{Markov chain Monte Carlo} \label{sec:mcmc}

A Markov chain Monte Carlo \cite{Gilks:1996} produces a set of samples
$\{ \vtheta^{(j)} \, | \, j = 1, \ldots \}$ from the model parameter
space that are sampled according to the posterior, meaning that, in
the limit that the chain length tends to infinity, the relative
frequency with which a given set of parameters appears in the chain is
proportional to the desired posterior, $p(\vtheta|d,M)$.  Therefore,
the output of an MCMC can be directly interpreted as the posterior PDF
over the full parameter space, while PDFs for individual parameters
can be obtained by marginalizing over the uninteresting parameters.

A Markov chain has the property that the probability distribution of
the next state can depend only on the current state, not on the past
history: 
\be 
p(\vtheta^{(j+1)})=\int_{V} d\vtheta^{(j)} p(\vtheta^{(j)} \to
\vtheta^{(j+1)}) p(\vtheta^{(j)}), \ee
where the jump probability $p(\vtheta^{(j)} \to \vtheta^{(j+1)})$
depends only on $\vtheta^{(j)}$ and $\vtheta^{(j+1)}$.  An additional
requirement for an MCMC arises from the fact that the desired
distribution is the equilibrium distribution.  Detailed balance requires that
$p(\vtheta^{(i)}) p(\vtheta^{(i)} \to \vtheta^{(j)}) = p(\vtheta^{(j)}) p(\vtheta^{(j)} \to \vtheta^{(i)})$.

% In other words, if we assume that state $(j)$ of the chain is sampled from the desired PDF, $p(\vtheta^{(j)})=p(\vtheta^{(j)}|d,M)$ , then the next state $(j+1)$ must be sampled from the PDF as well, so that $p(\vtheta^{(j+1)})=p(\vtheta^{(j+1)}|d,M)$; this condition is known as ``detailed balance.''

One way to produce such a sequence of samples is via the
Metropolis-Hastings algorithm, first proposed in
\cite{Metropolis:1953}, and later generalized in
\cite{Hastings:1970}:
\begin{enumerate}
\item Given a current state $\vtheta^{(j)}$, propose the next state
  $\vtheta^p$ by drawing from a jump proposal distribution with
  probability $Q(\vtheta^{(j)} \to \vtheta^p)$.
\item Compute the probability of accepting the proposed jump as
  \bel{eq:p-accept} 
p_{\textnormal{accept}} \equiv \min\Bigl(1,
\frac{p(\vtheta^p|d, M)}{p(\vtheta^{(j)}|d, M)} \frac{Q(\vtheta^p \to
  \vtheta^{(j)})}{Q(\vtheta^{(j)} \to \vtheta^p)} \Bigr).  
\ee
\item Pick a uniform random number $\alpha \in [0,1]$.  If $\alpha<
  p_{\textnormal{accept}}$, accept the proposed jump, setting
  $\vtheta^{(j+1)}=\vtheta^p$.  Otherwise, reject the jump, and remain
  at the same location in parameter space for the next step,
  $\vtheta^{(j+1)}=\vtheta^{(j)}$.
\end{enumerate}
 
This jump proposal distribution $Q(\vtheta^{(j)} \to \vtheta^p)$ can
depend on the parameters of the current state $\vtheta^{(j)}$, but not
on the past history.  It must also allow any state within the prior
volume to be reachable (eventually) by the MCMC.  Any jump proposal
that satisfies these properties is suitable for an MCMC.

The jump proposal is the most important choice in the MCMC, as it
determines the sampling efficiency of the algorithm, i.e., the length
of the chain before it converges to the posterior PDF.  Creating an
efficient jump proposal distribution requires an understanding of the
structure of the parameter space which may not be available until the
PDFs are found, creating a Catch-22; one possibility for resolving
this infinite loop is described in Section \ref{sec:higherdim}.

It should be noted that although an MCMC whose jump acceptance
criterium obeys detailed balance (as the Metropolis-Hastings algorithm
does) must eventually converge to the desired distribution, there is
no way to guarantee convergence in a fixed number of steps or to test
whether a chain has converged in a foolproof manner.  For example,
MCMC chains can get stuck on local maxima, producing an apparently
well-converged sampling of the PDF in the vicinity of the maximum; or,
if the chain visits a sequence of local maxima, moving rarely between
maxima, the autocorrelation length of the chain may represent a
substantial fraction of the total number of samples, resulting in an
effective sample size that is too small to accurately represent the
relative sizes of the modes in the PDF.

Finally, we note that, in practice, the randomly chosen initial
starting point of the MCMC may be in a particularly unlikely location
in the parameter space.  Because jumps are frequently local, we will
generally want to ignore the early samples in a finite-size chain to
avoid biases in the recovered posterior PDF due to the choice of the
initial location.  The samples thus discarded are referred to as
``burn-in'' samples. 

\subsection{RJMCMC}\label{RJMCMC}

The samples produced by an MCMC algorithm can be used to directly
perform a Monte Carlo evidence integral.  This results in a harmonic
mean estimator for the evidence, which may suffer from infinite variance \cite{NewtonRaftery:1994,Chib:1995,vanHaasteren:2009,WolpertSchmidler:2012}.
Additional techniques for the direct integration of evidence, also
based on a kD tree decomposition of the parameter space (see
Sec.~\ref{sec:kDTree}), are described in \cite{Weinberg2009}.  These
techniques are promising, but in some cases suffer 
from large variance and bias \cite{Farr2010}.  An alternative approach to model
selection among a set of models is based on performing an MCMC in a
``super-model'' that encompasses all of the models under
consideration; this is known the the Reversible Jump Markov chain
Monte Carlo (RJMCMC).

The parameter space of the super-model in an RJMCMC consists of a
discrete parameter that identifies the model, $M_i$, and a set of
continuous parameters appropriate for that model, $\vtheta_i$.  Thus,
each sample consists of a model identifier and a location within the
parameter space of that model, $\{M_i, \vtheta_i\}$.  We perform the
MCMC in the ``super-model'' parameter space just like a regular MCMC;
we propose jumps to different parameters within a model (intramodel
jumps) and jumps to a different model with different parameters
(intermodel jumps).  The acceptance probability for a proposed jump
from $\vtheta^{(j)}_i$ in model $M_i$ to $\vtheta^p_j$ in model $M_j$
becomes
\begin{equation}
p_{\textnormal{accept}} \equiv \min\Bigl(1,
\frac{p(\vtheta^p_j, M_j|d)}{p(\vtheta^{(j)}_i, M_i|d)}
\frac{Q(\vtheta^p_j ,M_j \to
  \vtheta^{(j)}_i, M_i)}{Q(\vtheta^{(j)}_i, M_i \to \vtheta^p_j, M_j)} \Bigr).
\end{equation}
Here the $Q$ factors incorporate both a discrete probability on the
model index, reflecting the probabilistic choice of which model to
jump into, and also a continuous probability density on target model's
parameter space.  For example,
$Q(\vtheta^p_j ,M_j \to \vtheta^{(j)}_i, M_i)$ has a factor for the
probability of proposing a jump to model $i$ when in model $j$, and is
a density on the parameter space of $M_i$.  These densities cancel the
corresponding densities in the ratio of posteriors, making the
acceptance probability a parameterisation-independent scalar.  In the
common special case where the two parameter spaces have equal
dimension and the jump proposal is a diffeomorphism, $\phi$, between
them,
\begin{equation}
  \vtheta^p_j = \phi\left( \vtheta^{(j)}_i \right),
\end{equation}
and therefore 
\begin{equation}
  \vtheta^{(j)}_i = \phi^{-1} \left( \vtheta^p_j \right),
\end{equation}
then the jump proposal ratio reduces to the Jacobian of the
diffeomorphism:
\begin{equation}
  \frac{Q(\vtheta^p_j ,M_j \to
  \vtheta^{(j)}_i, M_i)}{Q(\vtheta^{(j)}_i, M_i \to \vtheta^p_j, M_j)}
= \left| \frac{\partial \phi}{\partial \vtheta^{(j)}_i} \right|.
\end{equation}

 The resulting chain samples from the posterior
$p(M_i, \{\vtheta_i\}|d)$.  As in a usual MCMC, the PDF on the model as a
parameter, with other parameters ignored, is obtained by marginalizing
over the remaining parameters.  The posterior probability of a model
is proportional to the number of counts
\be
p(M_i|d) = \int d\vtheta_i \frac{L(d|M_i, \vtheta_i) p(\vtheta_i|M_i) p(M_i)}{p(d)}
\approx \frac{N_i}{N},
\ee
where $N_i$ is the number of RJMCMC samples listing the $i$'th model
and $N$ is the total chain length.  Thus, the probability of a
particular model relative to other models under consideration is given
by the fraction of RJMCMC samples lying in the parameter space of that
model.
 
The main difficulty of achieving an efficient RJMCMC is finding a good
jump proposal distribution for intermodel jumps.  In order to have
relatively high acceptance ratios for intermodel jumps, which is
necessary for efficient mixing between models, jumps should be
preferentially proposed into regions with a high posterior.  However,
because the algorithm is Markovian, it has no past memory, so a jump
proposed into a model from outside can not access information from
earlier in the chain which may identify a posterior peak.  It is, in principle, possible to overcome this constraint by storing a union of $\{\vtheta_i\}$ as the MCMC state vector, with the likelihood a function only of parameters $\vtheta_i$ that correspond to the current model $M_i$.  In this case, intermodel jump proposals would change only the model $M_i$.  However, if the chain finds a high-likelihood region in one model space faster than in another, a jump to the other model will take a very long time to be accepted -- again rendering RJMCMC inefficient.

The way to solve this problem is to identify a good jump proposal
distribution in advance, by exploiting information from single-model
MCMCs to generate efficient jump proposal distributions for our
reversible jump MCMC (Single-model MCMCs can take small local jumps
within their model, meaning that they are much less likely than an
RJMCMC to lose a high-posterior mode once it has been located).  The
ideal jump proposal distribution for the parameters within a model
would consist of the posterior PDF for those parameters,
$p(\vtheta_i|M_i,d)$, and single-model MCMCs already represent samples
from these posterior PDFs.  However, the samples are discrete, and a
jump proposal must be continuous.  Therefore, the output of each
single-model MCMC must be interpolated to construct the desired jump
proposal.  The strategy we propose for efficiently interpolating
a discretely sampled PDF is described in the next section.

\section{kD Trees and Interpolation}
\label{sec:kDTree}

The problem of drawing a proposed jump from an interpolation of
single-model MCMC samples can be thought of as the problem of assigning
a local ``neighborhood'' to each sample in the MCMC chain.
We choose these neighborhoods to be non-overlapping and to fill the parameter
space.  The size of a
neighborhood is inversely proportional to the local sample density.
The proposed jumps are drawn from a piecewise-constant (constant on
each neighborhood) interpolation of the PDF.  
To draw a proposed jump, we select a sample uniformly from the
MCMC samples, find its associated neighborhood, and then draw the
proposed jump uniformly from the neighborhood.  Since the MCMC samples
are distributed according to the posterior PDF for the single model,
this procedure produces proposed jumps that are approximately
distributed according to the posterior PDF.  

There are various
techniques that could be used to construct the set of neighborhoods
associated with each sample.
\cite{Littenberg2009} decompose the parameter space into
constant-volume ``bricks'' whose size is set by the typical size of
the peaks of the PDF.  Each sample is associated with the brick that
contains it, and the probability of proposing a jump into a particular brick is thus proportional to the
number of samples within that brick.  Additionally, an extra uniform jump proposal is added to allow for jumps into bricks that do not contain any samples, so that the jump 
proposal covers the entire model parameter space.  
However, the bricks in this algorithm do not adapt to the local
structure of the PDF.  One must either use small bricks to capture the
local structure of the PDF, placing many bricks in regions without
MCMC samples (which can increase memory management and access costs),
or use large bricks, missing the local structure of the PDF in
exchange for fewer empty bricks.

An alternate technique for producing adaptive neighborhoods would be
to use the Voronoi regions \cite{Voronoi1907} associated with each
MCMC sample.  The Voronoi region associated with a sample contains all
the parameter space points that are closer to that sample than any
other sample.  The Voronoi region decomposition into neighborhoods is,
in a sense, maximally adaptive, in contrast to the approach of
\cite{Littenberg2009}, which is minimally adaptive.
Unfortunately, defining the Voronoi decomposition requires a metric on
parameter space, which may be difficult or impossible to define.
Also, the computational cost for computing the Voronoi regions
increases rapidly with dimensionality.  
% Apparently the exact computational cost is not known in arbitrary dimension.

Here we propose to use a decomposition of the parameter space into
neighborhoods based on a data structure called a kD-tree (see,
e.g.\ \cite{Berg2008} or \cite{Gaede1998}).  The decomposition is more
adaptive than the boxes of \cite{Littenberg2009}, and more
efficient in high-dimensional spaces than the Voronoi decomposition.

\begin{figure}
  \begin{center}
    \includegraphics[width=0.8\columnwidth]{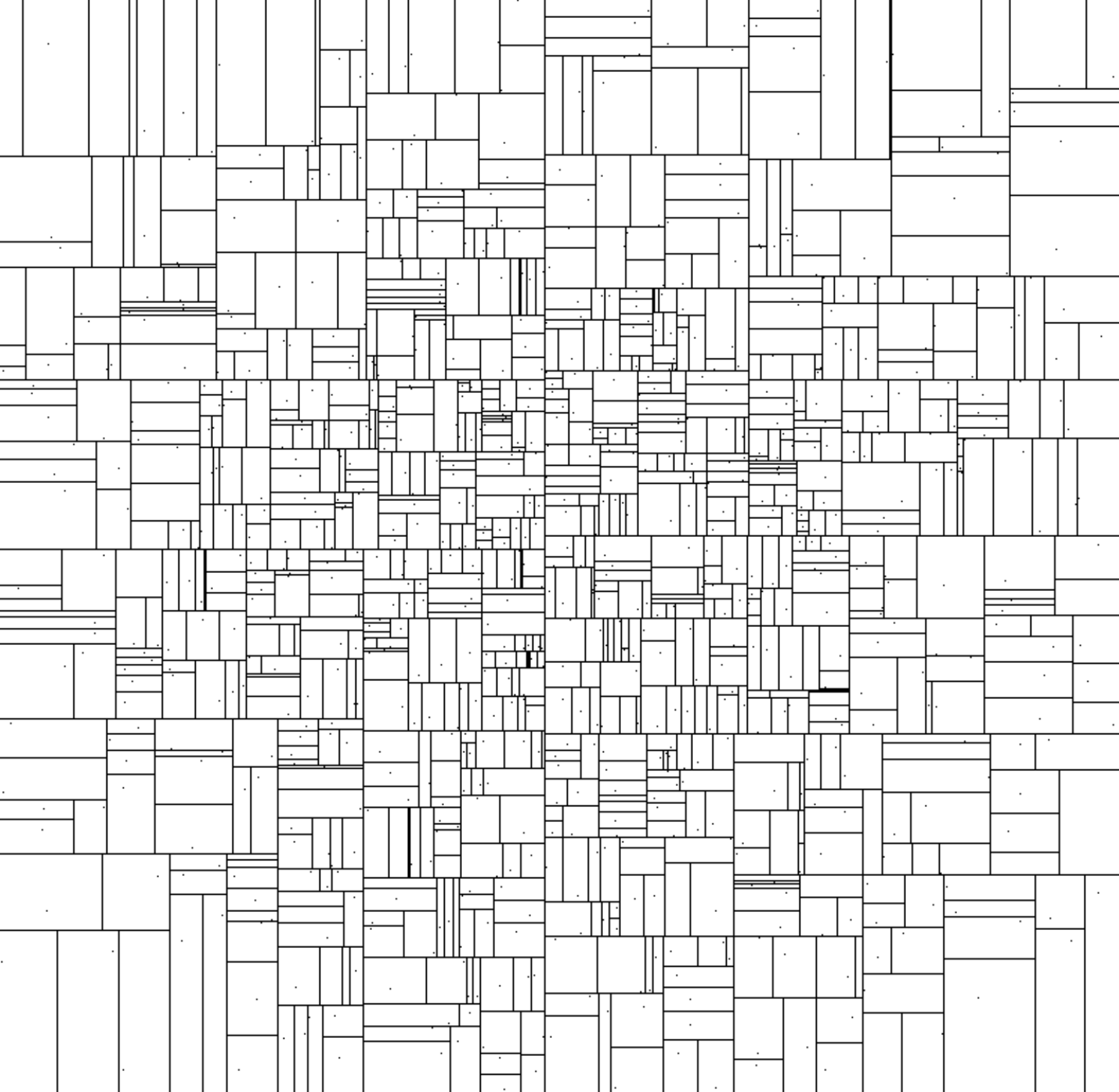}
  \end{center}
  \caption{\label{fig:kD-tree} The neighborhoods from a kD-tree
    constructed around a set of samples that are normally distributed
    about the origin in two dimensions.  As the samples become denser
    around the origin, the typical neighborhood gets smaller.  The
    interpolated PDF within a box of volume $V_i$ is $1/(N V_i)$,
    where $N$ is the total number of samples (which is also the number
    of boxes).}
\end{figure}

A kD-tree is a binary, space-partitioning tree.  To partition a set of
samples into a kD-tree, begin by placing them in a rectangular box that
contains all of parameter space.  Then proceed recursively%
\footnote{The kD-tree data structure defined here places box boundaries
  between the samples.  An alternate definition common in the
  literature places box boundaries on the median sample, but such a
  definition is inconvenient for our purposes.}: %
\begin{enumerate}
\item If the given box contains exactly one sample, stop; this is a
  leaf of the tree.  Otherwise:
\item Choose a dimension along which to divide the samples.  Divide the
  samples in half along this dimension (or nearly in half, if the
  number of samples is odd), forming two sub-boxes.  The ``left''
  sub-box contains the half (or nearly half) of the samples that have
  small coordinates along the chosen dimension; the ``right'' sub-box
  contains the half (or nearly half) of the samples that have large
  coordinates along the chosen dimension.
\item Return to Step 1 with each of the sub-boxes, storing the
  resulting trees as sub-trees of the current box.
\end{enumerate}
The key algorithmic step in the production of a kD-tree is finding the
median sample along a given dimension in order to divide the samples in
half in Step 2.  For $n$ samples, this can be accomplished in
$\order{n}$ time (see, e.g., \cite{Press2007}).  If there are $N$
samples in total, there are $\order{\log N}$ levels in the tree; at
each level, $\order{N}$ samples must be processed once in the
median-finding algorithm.  Tree construction thus costs $\order{N \log
  N}$ in time, and the tree consumes $\order{N}$ space.  As an
illustration, box boundaries for a kD-tree constructed around a sample
set that is normally distributed around the origin in two dimensions
are shown in Figure \ref{fig:kD-tree}.

In order to use the kD-tree interpolation as a jump proposal in an
MCMC, we randomly select a point stored in the kD tree with equal
probability, and propose from the associated neighbourhood.
Therefore, we must be able to quickly find the neighborhood associated
with a given point to compute the jump probability (see
Eq.~\ref{eq:p-accept}).  We introduce an additional parameter into the
neighbourhood search, $N_\mathrm{boxing}$, which describes the minimum
number of points in a kD box used as the neighbourhood for a point;
small $N_\mathrm{boxing}$ increases variance in the proposal but
resolves finer-scale structure in the posterior.  The neighbourhood
search can be accomplished in $\order{\log N}$ time and constant space
with the following algorithm, which is a modified binary tree search.
Given the randomly-chosen point, $\vtheta_i$, the tree, $T$, and
$N_\mathrm{boxing}$:
\begin{enumerate}
\item If $T$ contains fewer than $2N_\mathrm{boxing}$ points, then
  its box is the associated neighborhood.  Otherwise:
\item The tree $T$ has two sub-trees.  If the point $\vtheta_i$ is
  contained in the ``left'' sub-tree, then return to Step 1,
  considering this sub-tree; otherwise return to Step 1, considering
  the ``right'' sub-tree.
\end{enumerate}
The returned box is used for the jump proposal by drawing uniformly
from its interior, so the proposal density is 
\begin{equation}
  \label{eq:neighbourhood-jp}
  Q(\vtheta \to \vtheta^p) = \frac{N_\mathrm{box}}{N V},
\end{equation}
where $N$ is the number of points in the tree, $N_\mathrm{box}$ is the
number of points in the chosen kD box, and $V$ is the coordinate-space
volume of the box.

\section{RJMCMC Efficiency}
\label{sec:efficiency}

In this section, we demonstrate the efficiency of the kD-interpolated
jump proposal on a toy model-comparison problem.  The same algorithm
has been used in real-world settings \cite{Farr2010} and, as discussed
below, is the available in several forms as a software library for
re-use by others.

In the toy model of this section, we draw $N = 100$ simulated data
points from a $N(0,1)$ Gaussian distribution, and then ask whether
these data are better described by a model where they are Gaussian
distributed with unknown mean $\mu$ and standard deviation $\sigma$
\be
p(x) = \frac{1}{\sqrt{2\pi} \sigma} \exp\left( - \frac{(x-\mu)^2}{2
    \sigma^2} \right)\ ,
\ee
or by a model where they are Cauchy distributed with mode $\alpha$
and width $\beta$
\be
p(x) = \frac{1}{\pi \beta \left( 1 + \left(\frac{x - \alpha}{\beta}\right)^2\right)}\ .
\ee
We take priors on $\mu$ and $\alpha$ to be uniform in $[-1,1]$, and
priors in $\sigma$ and $\beta$ to be uniform in $[0.5, 1.5]$.  With a
data set of 100 points, the relative uncertainty in determining the
parameters of the underlying distribution is approximately 10\%, so we
expect the posterior probabilities in the $(\mu,\sigma)$ and
$(\alpha,\beta)$ spaces to occupy only a few percent of the prior
volume.  The Cauchy distribution is much broader than the Gaussian (it
has no finite moments), so with equal model priors, the posterior
probability for the Gaussian model over the Cauchy model is extremely
large:
\be
\frac{p(\textnormal{Gaussian} | d)}{p(\textnormal{Cauchy}|d)} \sim 10^9.
\ee
In order to ensure that the RJMCMC produces samples in the Cauchy
model at all, we impose a model prior that favors the Cauchy model by
$5 \times 10^8$ relative to the Gaussian.  The evidence ratio between
the models for our chosen data set with these priors is
\be \frac{p(\textnormal{Gaussian} | d)}{p(\textnormal{Cauchy}|d)}
\equiv r = 1.15,
\ee
yielding a theoretical maximum acceptance rate of inter-model jumps of
$(1+1/r)/2 = 0.93$.

\begin{figure}
  \begin{center}
    \includegraphics[width=0.8\columnwidth]{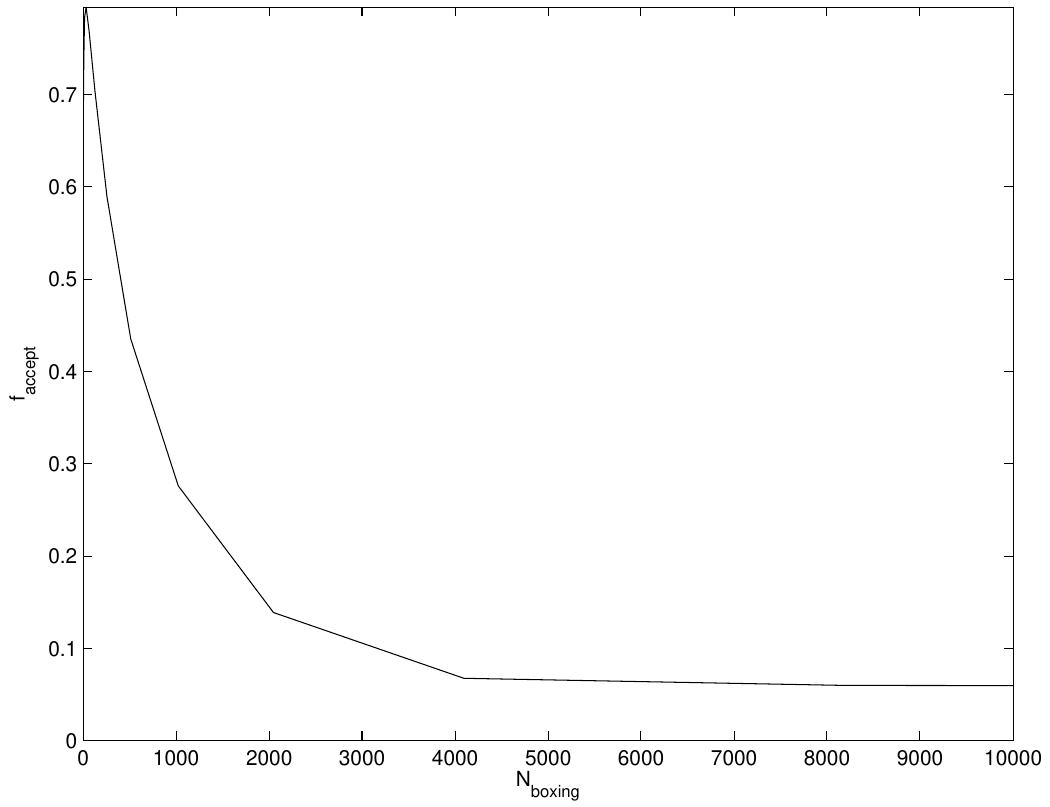}
  \end{center}
  \caption{\label{fig:acceptRate} The inter-model jump acceptance rate
    versus the number of samples per box when the kD-tree neighborhood
    search is truncated.  As the number of samples per box increases,
    and the interpolation becomes less accurate, the acceptance rate
    falls, asymptoting to the rate for naive draws from the uniform
    prior (about $5\%$ for this data set).}
\end{figure}

We obtain $10^4$ single-model MCMC samples by independently running
MCMC within each model, and use the kD-tree interpolation method
described above to propose inter-model jumps in an RJMCMC.  The
acceptance rate of inter-model jumps is approximately 0.8.  To explore
how the efficiency of the method degrades as the interpolation becomes
less accurate, we artificially truncated the kD tree with higher and
higher numbers of samples in each box (this can be accomplished during
the neighborhood search phase by stopping the search for a box when
one is found containing the desired number of samples).  For each
truncation choice, we performed an RJMCMC with the resulting
interpolated jump proposal.  The acceptance rate is plotted against
the number of single-model MCMC samples per box (kD-tree leaf) in
Figure \ref{fig:acceptRate}.  The more samples in each leaf of the tree
when the search is truncated, the lower the acceptance probability;
when points are drawn from the top level of the tree, the acceptance
probability asymptotes to the naive draw from the prior ($\sim 5\%$).

The relative error on the determination of the Bayes factor (evidence ratio) scales
with $1/\sqrt{N_{\textnormal{transitions}}}$, where
$N_{\textnormal{transitions}}$ is the number of inter-model
transitions in the RJMCMC.  Thus, as the acceptance rate of
inter-model jumps goes down, the RJMCMC must run longer to achieve a
desired accuracy in the evidence ratio.  By boosting the acceptance
rate of inter-model jumps, the interpolation method described above
can improve the runtime of an RJMCMC.

\section{kD-Interpolated Jump Proposal in Higher-Dimensional
  Single-Model MCMCs}
\label{sec:higherdim}

In the model selection example from Section \ref{sec:efficiency}, the
models have two-dimensional parameter spaces; in \cite{Farr2010} the
highest-dimensional model had five dimensions.  As the number of
dimensions increases the interpolation becomes more difficult for two
reasons.  First, the number of samples required for a given
number of subdivisions in each dimension grows exponentially with
the number of dimensions; hence, for a reasonable number of samples, a high-dimensional kD tree will have few subdivisions along each dimension.  Second, as
the dimensionality increases, the fraction of the volume at the ``edges'' and ``corners'' of each kD-tree box becomes more significant, and the placement of the sample becomes increasingly unrepresentative of the typical density in the box.  This problem is even more
pronounced when the distribution of samples does not align with the
coordinate axes along which the tree subdivides.  

In this section, we propose a practical solution that allows us to incorporate larger-scale features in the sample density distribution.  We illustrate our solution with an single-model MCMC that updates the jump proposal distribution on-the-fly by using a kD tree to interpolate the density of existing samples.  We conclude with an example of a successful application to gravitational-wave parameter estimation.

\begin{figure}
  \includegraphics[width=0.8\columnwidth]{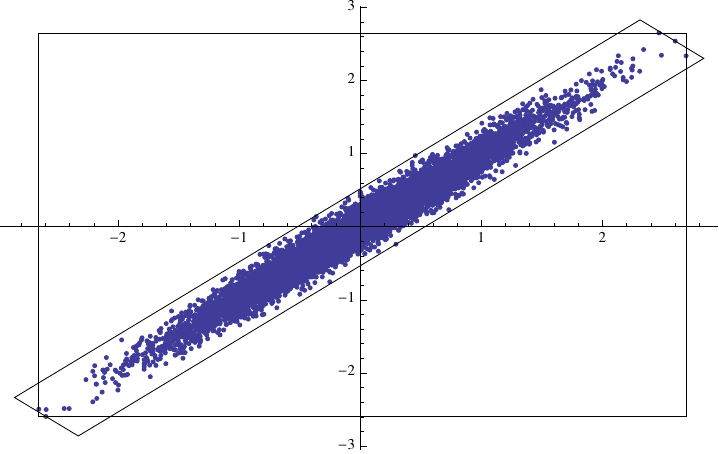}
  \caption{\label{fig:PCC} A two-dimensional illustration of the two
    boxes involved in the modified kD proposal.  The larger box
    aligned with the axes is the normal kD-tree box containing the
    given samples, while the tighter box is the ``covariance cell''
    aligned with the eigenvectors of the sample distribution.  The
    modified kD-interpolated jump proposal draws a point from the
    intersection of the two boxes. Tightly-correlated posteriors in
    parameter space such as this are typical of the gravitational-wave
    parameter estimation problem described in \protect\cite{Veitch:2014} and
    in the text.  Without the modification to account for the
    correlated samples, the kD neighbourhood of these points would
    produce a very inefficient jump proposal, since most of the
    bounding-box would contain empty (i.e., low-posterior) parameter
    space.}
\end{figure}

The kD-tree based interpolated jump proposal described in Section \ref{sec:kDTree} selects one sample from the tree at random and proposes a uniform point from the box containing that sample.  This ignores the density of other samples in the neighborhood of this box, which contains additional information about the true density within the box, which is likely non-uniform.  A better proposal, therefore, would account for larger-scale sample covariances in parameter space.  To address significant covariance between samples, which would correspond to strong density gradients and a very non-uniform density profile within a rectangular box, the modified jump proposal does not descend to the leaf-nodes of the tree, but instead stops descending whenever the number of samples in the current box, $\Nbox$, falls below some threshold, $\Ncrit$. We then find the principal axes of the samples
contained in this box by calculating the eigenvectors of the covariance matrix of these samples. We use these eigenvectors to draw a new box. This box -- henceforth called a covariance cell -- contains all $\Nbox$ samples from the kD-tree box, is centered on the mean of these samples, and has edges that are aligned with the principal axes. Furthermore, this box is drawn around the samples as tightly as possible; in other words, for each edge, there exists a point in the covariance cell that lies along that edge. Figure \ref{fig:PCC} illustrates these two boxes.

The jump proposal chooses a stored sample at random and finds the largest kD-tree box containing it and fewer than $\Ncrit$ total samples. It then draws the covariance cell and proposes a new point uniformly from the intersection of the covariance cell with the original kD-tree box.
The jump probability for this jump proposal is still (c.f.\ Eq.\
\eqref{eq:neighbourhood-jp}) given by
\bel{modforward} Q(\vtheta_i \rightarrow \vtheta_{i+1}) =
\frac{\Nbox}{N V} , \ee
where $N$ is the total number of samples in the kD tree, and $V$ is the
volume of the intersection between the kD-tree box and the correlation cell containing $\vtheta_{i+1}$. When used as the only jump proposal for an
RJMCMC, it is important that this proposal be capable of
proposing points in all allowed regions of parameter space; on the
other hand, when the kD proposal is used in a suite of other jump proposals, this is no longer a requirement.

%The backward jump probability is computed in a similar fashion by
%finding the principal-component box that would contain the current set
%of parameters: \ilya{[Probably not necessary -- we don't explicitly provide backward jump proposals in earlier sections]}
%
%\bel{modbackward} P(\vtheta_{i+1} \rightarrow \vtheta_{i}) =
%\frac{\Nbox'}{N\left(V' \cap \Vbox'\right)} , \ee
%
%where $\Nbox'$ is the number of points in the original box containing
%$\vtheta_i$, $V'$ is the volume of the kD box containing $\vtheta_i$,
%and $\Vbox'$ is the volume of the principal-axes box containing
%$\vtheta_i$.  Note, however, that the current parameters may lie in a
%region of the kD box that does not intersect the principal axes box;
%in this case, the backward jump probability is zero, and the algorithm
%rejects the proposed step.

This technique trivially violates detailed balance since the jump proposal distribution depends on the past history. One way to address this problem, through diminishing adapations \cite{Brooks:2011}, requires continuously accumulating samples into the kD tree as the MCMC explores more
of the parameter space. As the number of samples drawn from the posterior PDF increases, the neighborhoods around each sample decrease with increasing sample density and the calculated covariances between samples in the kD tree better approximate the true covariances between parameters in the equilibrium posterior PDF.  Hence, for large $N$, the change in $Q(\vtheta^{(j)} \rightarrow \vtheta^{(j+1)})$ approaches zero and the jump proposal does a progressively better job of sampling the equilibrium posterior while becoming asymptotically Markovian.  Conservatively, the entire set of samples obtained while the jump proposal is being dynamically updated could be discarded as burn-in, and future analysis could use the static jump proposal distribution interpolated from samples accumulated during the extended burn-in phase. 
%Because this introduces a dependence on the
%previous states of the MCMC, the jump proposal is no longer
%Markovian.  However, it remains \emph{asymptotically} Markovian \ilya{[And being asymptotically Markovian (whatever that actually means -- at least a reference?) -- is provably sufficient to guarantee that we sample from the equilibrium distribution even though we nominally violate detailed balance? (again, a reference?)]} , which
%ensures that the equilibrium distribution of the MCMC is correct.

To efficiently insert samples into the tree as the chain accumulates them,
the algorithm from Section \ref{sec:kDTree} must be modified:
\begin{enumerate}
\item Instead of partitioning the sample set at its median, we now
  partition the bounding box at its geometric center along a
  particular dimension, which cycles as we descend the tree levels.
  Note that this allows for empty boxes if all the samples cluster to
  one side of the domain.
\item When inserting a sample, we descend the tree to a leaf node,
  which now can be either empty or contain one sample.  If it is empty,
  we insert the sample at this node.  If it contains one sample, we
  subdivide the appropriate dimension, and place both samples into the
  two sub-boxes.  If both land in the same sub-box, we continue
  subdividing, until each box in the sub-sub-sub... tree contains one
  or zero samples.
\end{enumerate}
%Insertion of a sample is $\order{\log N}$, since it involves a constant amount of work per tree level, and there are $\log N$ tree levels.  \ilya{[No -- the number of tree levels can be arbitrarily large if the original samples are clustered into a very small region of the prior volume...]} \dan{[Is this true? Couldn't it be arbitrarily small, depending on how far into the tree you want to burrow (i.e. depending on $N_{crit}$)?]} Neighborhood lookup proceeds as before, and is therefore also $\order{\log N}$.  \ilya{[Perhaps $\order{\log N + \Nbox}$, since must recompile the inner bounding box every time -- and, depending on problem, $\Nbox$ could be $\gg \log N$?]} \dan{[Need only recompute the covariance cell if a new sample was accepted into the corresponding kD tree cell between calls to that kD tree cell. I'm not sure how that would factor in here, but I feel like it should.] }

We have implemented this modified kD tree proposal as one of many jump
proposals in the \texttt{LALInferenceMCMC} sampler
\cite{vanderSluys:2008a,Raymond:2010,Raymond2012,Veitch:2014}.
\texttt{LALInferenceMCMC} is a MCMC code, based on the LIGO algorithms
library ({\tt http://www.lsc-group.phys.uwm.edu/lal}), designed to sample the posterior on parameters of merging
compact-object binaries (masses, sky location, orbital orientation, distance, etc.) 
encoded in their gravitational-wave signatures as observed by the ground-based
gravitational-wave detectors LIGO \cite{AdvLIGO} and Virgo \cite{AdvVirgo}.  The simplest such
signal has a nine-dimensional parameter space, and the posterior often
includes near-degeneracies and multiple modes that make convergence of
the MCMC chain very slow with traditional proposals \cite{S6PE}.  

\begin{figure}
  \begin{center}
    \includegraphics[width=0.8\columnwidth]{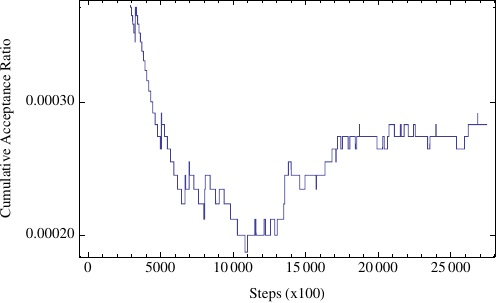}
  \end{center}
  \caption{\label{fig:accratio} The cumulative acceptance
    ratio for the modified kD-tree jump proposal used in the
    \texttt{LALInferenceMCMC} code as a function of the number of
    steps (in hundreds). The simulation in question was a
    nine-dimensional analysis of a simulated gravitational-wave signal
    injected into synthetic data similar to that taken by the LIGO
    and Virgo detectors \protect\cite{AdvLIGO,AdvVirgo}.  The parameter space
    includes the masses of the compact objects generating the
    gravitational-wave, their location on the sky, distance, orbital
    orientation, the time of signal arrival, and the orbital phase.
    The posterior in this problem has a number of well-separated modes
    in the parameter space which are difficult to jump between using
    traditional jump proposals; in spite of the small acceptance ratio
    of the kD proposal, when applied to one in twenty jumps proposed
    in this simulation, it improved the convergence time of the sampler by a factor of two
    compared to using only the standard suite of proposals. 
    The acceptance rate asymptotes to the steady-state solution once sufficient samples have been accumulated in the kD tree to allow the sample density to be accurately interpolated; samples collected
prior to this point should be discarded as burn-in.
  }
\end{figure}

Figure \ref{fig:accratio} shows the acceptance ratio for the kD-tree
jump proposal applied as part of a \texttt{LALInferenceMCMC} analysis.
In this example, the kD-tree jump proposal is one of several jump
proposals used during the MCMC, and the kD tree itself is updated with
accepted samples whenever its jump proposal is called. In spite of the
very low acceptance rate of the proposal, applying the kD-tree
proposal to one out of 20 proposed jumps improved the convergence
time---defined as the average time to reach a specified number of
independent samples from the posterior, thinning by the
autocorrelation length as described in \cite{Veitch:2014}---of the
simulation by a factor of two compared to the standard suite of
proposals because it is particularly efficient at producing
``mode-hopping'' jumps, which are difficult to produce with the other
proposals in \texttt{LALInferenceMCMC}.

\section{Conclusion}
\label{sec:conclusion}

The need to compare evidences for multiple models arises in a large
variety of physical and astronomical contexts.  In this paper, we
described a technique that allows for efficient evidence
computations via a Reversible-Jump Markov chain Monte Carlo.  This
technique solves the usual problem of finding good inter-model jump
proposals in an RJMCMC by using a kD tree to quickly and accurately
interpolate an approximate posterior PDF from a single-model MCMC run,
and then proposing efficient inter-model
jumps from this interpolated PDF.

We demonstrated the efficiency of this technique on a toy
model-comparison problem described in Section \ref{sec:efficiency}.
We also successfully applied this technique to the problem of
selecting the best model for the observed distribution of black-hole
X-ray binary masses, as described in \cite{Farr2010}.  In addition to
model comparison, the PDF interpolation described here can be useful
in single-model MCMCs to inform the jump proposal distribution on the fly in order to propose jumps that can efficiently sample the parameter space (see Section \ref{sec:higherdim}), or to test MCMC convergence. 

We have made our implementation of the technique described in this
paper publicly available online at
\url{http://github.com/farr/mcmc-ocaml}, and also in the
\texttt{LALInferenceMCMC} sampler, at
\url{https://www.lsc-group.phys.uwm.edu/daswg/projects/lalsuite.html}.
We welcome readers to take advantage of this toolkit.

\section*{Acknowledgments}

We are grateful to Neil Cornish for interesting discussions.  WF acknowledges support
from NSF grant AST0908930 and NASA grant NNX09AJ56G.  IM
acknowledges support from the NSF Astronomy and Astrophysics
Postdoctoral Fellowship, award AST-0901985, for the early stages of this work.  
 DS acknowledges
support from a Northwestern University Summer Undergraduate Research Grant. IM is grateful for the hospitality of the Monash Center for Astrophysics supported by a Monash Research Acceleration Grant (PI Y.~Levin).  This work has been partially supported by the UK Science and Technology Facilities Council.

\section*{}

\bibliography{paper}

\end{document}